\documentclass[11pt,a4paper]{article} 
\usepackage{jheppub}

\usepackage{latexsym}
\usepackage{amsmath,amssymb}
\usepackage{graphicx}
\usepackage{subfigure}
\usepackage{xcolor}

\title{Hawking-Page phase transition in BTZ black hole revisited} 

\author[a]{Myungseok Eune,} %
\author[a,b,c]{Wontae Kim,} %
\author[b]{and Sang-Heon Yi} %
                                           
\affiliation[a]{Research Institute for Basic Science, Sogang
  University, Seoul, 121-742, Republic of Korea} %
\affiliation[b]{Center for Quantum Spacetime, Sogang University, Seoul
  121-742, Republic of Korea} %
\affiliation[c]{Department of Physics, Sogang University, Seoul
  121-742, Republic of Korea} %
                                           
\emailAdd{younms@sogang.ac.kr} %
\emailAdd{wtkim@sogang.ac.kr} %
\emailAdd{shyi@sogang.ac.kr} %
                                        
\abstract{We consider the Hawking-Page phase transition between the
  BTZ black hole of $M \ge 0$ and the thermal soliton of $M=-1$. In
  this system, there exists a mass gap so that there does not seem to
  exist a continuous thermodynamic phase transition.  We consistently
  construct the off-shell free energies of the black hole and the
  soliton by properly taking into account the conical space. And then,
  the continuous off-shell free energy to describe tunneling effect
  can be realized through non-equilibrium solitons.}

\keywords{Black Hole, Thermodynamics}

\begin{document}

\maketitle

\section{Introduction}
\label{sec:intro}

Since it has been claimed that a black hole has an
entropy~\cite{Bekenstein:1973ur}, thermodynamics of black holes has
been one of the most important arena in black hole
physics~\cite{Gross:1982cv,Hawking:1982dh,York:1986it}.  And then,
there have been extensive studies for thermodynamics and phase
transitions in various black holes~\cite{Myers:1988ze, Banados:1992wn,
  Mann:1990ci, Nappi:1992as, Zaslavsky:1994dx, Reznik:1994py,
  Peca:1998cs, Nojiri:1999pm, Caldarelli:1999xj, Medved:2000fk,
  Nojiri:2001aj, Cvetic:2001bk, Cai:2006td,Anninos:2008sj,
  Cadoni:2009bn, Banerjee:2010sd, Banerjee:2011au}.  For instance, in
Einstein gravity, the hot flat space in the cavity is more probable
than the large black hole below a critical temperature while the large
black hole is more probable than the hot flat space above the critical
temperature~\cite{York:1986it}.  Such a phase transition can be read
off from the on-shell free energy and the heat capacity.

In particular, for the Ba\~nados-Teitelboim-Zanelli (BTZ) blak hole
system~\cite{Banados:1992wn}, there are two distinct solutions of the
BTZ black hole of $ M \ge 0$ and the thermal soliton of the global
AdS${}_3$ whose mass is $M=-1$~\cite{Horowitz:1998ha, Surya:2001vj,
  Kleban:2004rx}.  The Hawking-Page (HP) phase transition has been
intensively studied by using the on-shell and off-shell free energies
as well as conical singularity considerations in
Refs.~\cite{Carlip:1993sa, Mann:1996ze, David:1999zb, Kurita:2004yn,
  Myung:2005ee, Myung:2006sq, Maloney:2007ud, Ogawa:2011fw,
  Grumiller:2012rt}.
Recently, the Euclidean path integral has been calculated by taking
into account the conical ensemble~\cite{Grumiller:2012rt}, where the
metrics are not differential but continuous at the horizon because of
the Riemannian geometry with the conical
singularity~\cite{Fursaev:1995ef}.

Generically, it is possible to elaborate thermodynamic evolutions by
the use of the off-shell free energy which plays a role of the
potential at the given temperature of heat
reservoir~\cite{Hawking:1982dh, York:1986it}.  The black hole below
the critical temperature decays to the thermal vacuum while the
thermal vacuum tends to tunnel to the black hole above the critical
temperature. The tunneling effect in the phase transition is crucial
since the black hole state can not avoid the barrier of the off-shell
free energy in order to arrive at the vacuum state and vice versa. The
extrema of the off-shell free energy are equilibrium states, and the
others correspond to non-equilibrium states.  All these states are
connected with each other so that the off-shell free energy should be
continuous everywhere. However, in the BTZ black hole, the mass
spectrum is discontinuous so that at first glance the continuous
evolution seems to be impossible in the presence of the mass gap.

In this work, we would like to reconsider the HP phase transition
between the BTZ black hole and the thermal soliton by using the
off-shell free energy.  In fact, it would be interesting to answer how
to interpolate the states between the BTZ black hole $(M \ge 0)$ and
the thermal soliton $(M=-1)$ in order to elaborate the tunneling
effect which is essential in the HP phase transition.  For this
purpose, we will construct the off-shell free energy so that the black
hole state and the soliton state can be interpolated continuously.

In section~\ref{sec:thermodynamics:btz}, we shall recapitulate the
on-shell free energy which is just the free energy without considering
conical singularities.  In other words, one can define the free energy
from the Euclidean path-integral, and it turns out to be the on-shell
if the conical singularity contributions are neglected.  As expected,
the on-shell free energy describes equilibrium states of the black
hole and the soliton.  Next, in section~\ref{sec:HP_transition}, in
order to construct the off-shell free energy, we will calculate the
free energy of the black hole by taking into account the conical
singularity at the event horizon. Then, it can be shown that it is the
same with the off-shell free energy constructed from the conventional
definition of $F^{\rm{off}}_{\rm bh} = E - TS$, where $E$ and $S$ are
the thermodynamic energy and the entropy, and $T$ is an arbitrary
temperature. Thus, the free energy reflecting the conical singularity
can be identified with the off-shell free energy.  Of course, the
on-shell points can be obtained from the extrema of the off-shell free
energy, where it amounts to removing the conical singularity of the
metric.  Then, we will apply this notion to the side of the soliton in
such a way that we can obtain the off-shell states by taking into
account the conical singularity arising from the soliton.  After all,
the continuous off-shell free energy to describe tunneling effect can
be realized through non-equilibrium solitons.  Finally, summary and
discussion are given in section~\ref{sec:discus}.

\section{On-shell free energy}
\label{sec:thermodynamics:btz}

Let us begin with the Einstein-Hilbert action with a negative
cosmological constant in the three-dimensional spacetime
$\mathcal{M}$, which is given by
\begin{align}
  I_{\rm g} = \frac{1}{16\pi G} \int_{\mathcal{M}} d^3 x \sqrt{-g}
  (R-2\Lambda), \label{I:G}
\end{align}
where $G$ is the gravitational constant, and the cosmological constant
$\Lambda$ is related to the radius $\ell$ by $\Lambda \equiv
-1/\ell^2$. The Gibbons-Hawking term and the counterterm on the
boundary $\partial \mathcal{M}$ are written
as~\cite{Banados:1992wn,Henningson:1998gx,Balasubramanian:1999re}
\begin{align}
  I_{\rm GH} &= - \frac{1}{8\pi G} \int_{\partial \mathcal{M}} d^2 x
  \sqrt{|\gamma|} K, \label{I:GH} \\
  I_{\rm ct} &= - \frac{1}{8\pi G \ell} \int_{\partial \mathcal{M}}
  d^2 x \sqrt{|\gamma|}, \label{I:ct}
\end{align}
where $\gamma_{ij}$ and $K$ are the induced metric and the extrinsic
curvature scalar on the boundary $\partial \mathcal{M}$, respectively.
Then, the total action is given by
\begin{align}
  I = I_{\rm g} + I_{\rm GH} + I_{\rm ct}. \label{I:tot}
\end{align}
Varying the action~\eqref{I:tot}, we obtain the equations of motion as
\begin{align}
  R_{\mu\nu} - \frac12 g_{\mu\nu} R - \frac{1}{\ell^2} g_{\mu\nu} = 0. \label{eom}
\end{align}
The nonrotating BTZ solution to satisfy Eq.~\eqref{eom} is written
as~\cite{Banados:1992wn}
\begin{align}
  ds^2 = - f(r) dt^2 + \frac{dr^2}{f(r)} + r^2 d\phi^2, \label{metric}
\end{align}
where $ f(r) = -M + \frac{r^2}{\ell^2}$. The line
element~\eqref{metric} describes the geometry of the BTZ black hole,
and the event horizon is well defined at $r_H = \ell \sqrt{M}$.  The
integration constant $M$ is the ADM mass which is assumed to be $M>0$.
Moreover, the Hawking temperature and the Bekenstein-Hawking entropy
of the BTZ black hole are given by
\begin{align}
  T &= \frac{\kappa}{2\pi}=\frac{\sqrt{M}}{2\pi \ell}, \label{T:H}\\
  S_{\rm BH} &=\frac{A_H}{4G} = \frac{\pi \ell
    \sqrt{M}}{2G}, \label{entropy:BH}
\end{align}
where $\kappa$ is the surface gravity and $A_H = 2\pi r_H = 2\pi \ell
\sqrt{M}$.  Note that the geometry governed by Eq.~\eqref{metric}
becomes the AdS soliton which is just the global AdS${}_3$ spacetime
if $M=-1$~\cite{Banados:1992wn, Horowitz:1998ha, Hohm:2010jc,
  Perez:2011qp}.

In order to study thermodynamics of the system, we need to consider
the Euclidean geometry, and thus from now on all quantities will be
given in the Euclidean ones. The line element~\eqref{metric} becomes
\begin{align}
  ds_{\rm E}^2 = f(r) d\tau^2 + \frac{dr^2}{f(r)} + r^2
  d\phi^2, \label{metric:euclidean}
\end{align}
using $\tau = it$. Then, the action~\eqref{I:G}, the Gibbons-Hawking
boundary term~\eqref{I:GH}, and the counterterm~\eqref{I:ct} become
\begin{align}
  I_{\rm g} &= - \frac{1}{16\pi G} \int_{\mathcal{M}} d^3 x
  \sqrt{g} (R-2\Lambda), \label{I:G:E} \\
  I_{\rm GH} &= \frac{1}{8\pi G} \int_{\partial \mathcal{M}}
  d^2 x \sqrt{\gamma} K, \label{I:GH:E} \\
  I_{\rm ct} &= \frac{1}{8\pi G \ell} \int_{\partial \mathcal{M}} d^2
  x \sqrt{\gamma}, \label{I:ct:E}
\end{align}
where the boundary $\partial \mathcal{M}$ is the hyperspace given by
the constant $r$. For the first case of the BTZ black hole with $M>0$,
the actions are calculated as
\begin{align}
  I_{\rm g} &= \frac{\beta}{4G} \left( \frac{r^2}{\ell^2}
    - \frac{r_H^2}{\ell^2} \right),  \label{I:G:E:val} \\
  I_{\rm GH} &= \frac{\beta}{4G} \left( M -
    \frac{2r^2}{\ell^2} \right), \label{I:GH:E:val} \\
  I_{\rm ct} &= \frac{\beta r}{4G \ell}
  \sqrt{f}, \label{I:ct:E:val}
\end{align}
respectively. The total Euclidean action for the BTZ black hole
becomes
\begin{align}
  I_{\rm bh} &= I_{\rm g} + I_{\rm GH} + I_{\rm ct} \notag \\
  &= \frac{\beta r}{4G \ell} \left(\sqrt{f} - \frac{r}{\ell} \right) \notag \\
  &= - \beta \left[ \frac{M}{8G} + O\left(\frac{1}{r^2}\right)
  \right]. \label{I:E:val}
\end{align}
We now take the period of Euclidean time $\beta$ as the inverse of the
Hawking temperature~\eqref{T:H}, then $ I_{\rm bh} = I_{\rm bh}^{\rm
  on}$.  At the infinite boundary of $r \to \infty$, the thermodynamic
energy, the entropy, and the free energy are obtained as
\begin{align}
  E &= \frac{\partial I_{\rm bh}^{\rm on}}{\partial \beta} =
  \frac{M}{8G}, \label{energy:BTZ} \\
  S &= \beta E - I_{\rm bh}^{\rm on} = \frac{\pi \ell
    \sqrt{M}}{2G}, \label{entropy:BTZ} \\
  F_{\rm bh}^{\rm on } &= \beta^{-1} I_{\rm bh}^{\rm on} = -
  \frac{M}{8G}, \label{F:BTZ}
\end{align}
where $A_H = 2\pi \ell \sqrt{M}$ is the area of the horizon. Note that
the entropy~\eqref{entropy:BTZ} agrees with the Bekenstein-Hawking
entropy~\eqref{entropy:BH}. The heat capacity is given by
\begin{align}
  C = \frac{\partial E}{\partial T} = \frac{\pi \ell
    \sqrt{M}}{2G}, \label{C:BTZ}
\end{align}
where the BTZ black hole is stable since the heat capacity is always
positive.  Note that we have considered thermodynamic quantities
obtained form the Euclidean metric without conical singularities.

Next, for the soliton of $M=-1$, the actions are calculated as
\begin{align}
  I_{\rm g} &= \frac{\beta}{4G} \frac{r^2}{\ell^2}, \label{I:G:E:AdS}
  \\
  I_{\rm GH} &= \frac{\beta}{4G} \left( -1 - \frac{2r^2}{\ell^2}
  \right), \label{I:GH:E:AdS} \\
  I_{\rm ct} &= \frac{\beta}{4G} \frac{r}{\ell} \sqrt{1 +
    \frac{r^2}{\ell^2}}, \label{I:ct:E:AdS}
\end{align}
where the period of the Euclidean time has been chosen as $\beta$, so
that the total action of the AdS soliton is obtained as
\begin{align}
  I_{\rm sol}^{\rm on} &= I_{\rm g} + I_{\rm GH} + I_{\rm ct} \notag \\
  &= - \frac{\beta}{8G} \left[1 + O\left(\frac{1}{r^2} \right)
  \right]. \label{I:E:AdS}
\end{align}
With the infinite boundary, the free energy of the thermal soliton is
given by
\begin{align}
  F_{\rm sol}^{\rm on} = \beta^{-1} I_{\rm sol}^{\rm on} =
  -\frac{1}{8G}. \label{F:AdS}
\end{align}
Actually, the temperature $T=\beta^{-1}$ is the temperature of the
heat reservoir, and the soliton has the same temperature with that of
the reservoir.
\begin{figure}[htb]
  \centering
  \includegraphics[width=0.6\textwidth]{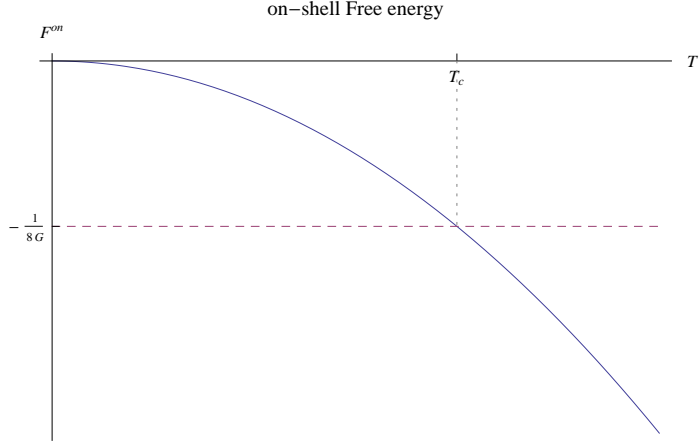}
  \caption{The solid and the dashed lines show the on-shell free
    energies of the BTZ black hole and the thermal soliton,
    respectively.}
  \label{fig:FT}
\end{figure}

The free energies with respect to the temperature are shown in
Fig.~\ref{fig:FT}, which shows that the phase transition occurs at
$T_c = 1/(2\pi\ell)$. As is well-known, the thermal soliton is more
probable than the black hole below the critical temperature while the
black hole is more probable than the thermal soliton above the
critical temperature.

\begin{figure}[htb]
  \centering
  \includegraphics[width=0.6\textwidth]{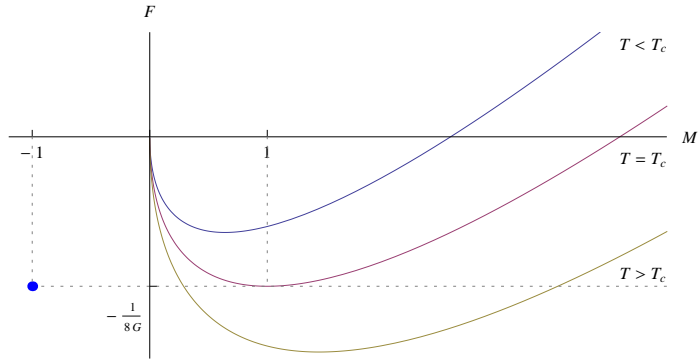}
  \caption{The solid lines show the off-shell free energies of the BTZ
    black hole depending on given temperatures, and the thick dot at
    $M=-1$ is the on-shell free energy of the thermal soliton. There
    are some missing off-shell states for $M <0$.}
  \label{fig:FM}
\end{figure}
\section{Off-shell free energy}
\label{sec:HP_transition}

Let us first consider the space $\hat{\mathcal{M}}$ with a conical
singularity, and denote the singular set by $\Sigma$ and the conical
deficit angle by $\Delta_\Sigma$. Then, the Ricci tensor and the
curvature scalar in the conical space are given
by~\cite{Fursaev:1995ef}
\begin{align}
  \hat{R}_{\mu\nu} &= R_{\mu\nu} + (n_\mu n_\nu) \Delta_\Sigma
  \delta_\Sigma, \label{Ricci:con} \\
  \hat{R} &= R + 2 \Delta_\Sigma \delta_\Sigma, \label{R:con}
\end{align}
where $\delta_\Sigma$ is the delta function defined by
$\int_{\hat{\mathcal{M}}} f \delta_\Sigma = \int_\Sigma f$ for any
function $f$ and $n^k = n^k_\mu dx^\mu$ ($k=1,2$) are two unit vectors
orthogonal to $\Sigma$, $(n_\mu n_\nu) = \delta_{ij} n^i_\mu n^j_\nu$.
And $R_{\mu\nu}$ and $R$ are the Ricci tensor and the curvature scalar
calculated in the regular points $\mathcal{M}/\Sigma$ by the standard
method, respectively.  From Eq.~\eqref{R:con}, the integration of the
curvature scalar $\hat{R}$ for the conical space is related to the
curvature scalar $R$ at the ordinary space as follows
\begin{align}
  \label{R:coni.sing}
  \int_{\hat{\mathcal{M}}} \sqrt{g} \hat{R} =
  \int_{\hat{\mathcal{M}}/\Sigma} \sqrt{g} R + 2 \Delta_\Sigma
  A_\Sigma,
\end{align}
where $A_\Sigma = \int_\Sigma$ is the area of the space $\Sigma$ for
the fixed cone.

For the case of the BTZ black hole of $M>0$, the conical singularity
appears from the Euclidean time at the event horizon. For the
arbitrary period of $\beta$, the conical deficit angle and the area of
$\Sigma$ are calculated as
\begin{align}
  \Delta_\Sigma &= 2\pi - \beta r_H /\ell^2 = 2\pi - \beta
  \sqrt{M} /\ell, \label{delta:Sigma:bh} \\
  A_\Sigma &= 2\pi r_H = 2\pi \ell \sqrt{M}. \label{A:Sigma:bh}
\end{align} 
Then, the total action~\eqref{I:tot} implemented by
Eq.~\eqref{R:coni.sing} is written as
\begin{align}
  I^{\text{off}}_{\rm bh} &= I_{\text{bh}}+I_{\text{sing}}  \notag \\
  &=\frac{\beta M}{8G} - \frac{\pi \ell
    \sqrt{M}}{2G}, \label{I:BH:off}
\end{align}
where $I_{\rm sing} = \beta M/4G - \pi \ell \sqrt{M} / (2G)$ from the
second term in Eq.~\eqref{R:coni.sing}.  Thus, we can define the
off-shell action of the black hole which consists of the regular
contribution from Eq.~\eqref{I:E:val} and the singular contribution
from Eq.~\eqref{R:coni.sing}.  Then it is natural to obtain the
off-shell free energy as
\begin{equation}
  \label{F:BH:off}
  F^{\rm off}_{\rm bh} = \beta^{-1} I^{\rm off}_{\rm bh} =
  \frac{M}{8G} - \frac{\pi \ell \sqrt{M}}{2G \beta}. 
\end{equation}
In this section, we define $\beta$ as the arbitrary temperature of the
heat reservoir, and $\beta_{\text{H}}$ as the inverse of the Hawking
temperature presented in Eq.~\eqref{T:H} for convenience.  Note that
the off-shell free energy~\eqref{F:BH:off} is reduced to the on-shell
free energy~\eqref{F:BTZ}, $\left.F^{\rm off}_{\rm bh}\right|_{\beta =
  \beta_H} = -M/8G$ by identifying $\beta=\beta_{\text{H}}$.  This
condition amounts to taking extrema of the off-shell free energy since
the extrema can be guaranteed by the thermodynamic first law which
satisfies $\beta=\beta_{\text{H}}$.  However, for the case of
$\beta\neq\beta_{\text{H}}$, the free energy is still in the
off-shell, which means that the thermodynamic system is in
non-equilibrium. It is interesting to note that the off-shell free
energy~\eqref{F:BH:off} can be written in the form of the well-known
Helmholtz free energy,
\begin{align}
  F^{\rm{off}}_{\rm bh} = E - TS, \label{Helmholtz}
\end{align}
once we identify the on-shell quantities of the energy $E$ and the
entropy $S$ with Eqs.~\eqref{energy:BTZ} and~\eqref{entropy:BTZ},
respectively while the temperature $T$ is treated as a variable.

In connection with the conical singularity, we would like to mention
compatibility of off-shell structures with the Einstein
equation~\eqref{eom} in Euclidean gravity.  For this purpose, let us
first consider Eq.~\eqref{eom} in the absence of any conical
singularities, then it gives on-shell solutions corresponding to
classical trajectories.  However, if we allow the conical singularity
in space, then the equation of motion should be replaced by
\begin{equation}
  \label{eom:modified}
  \hat{R}_{\mu\nu} - \frac12 g_{\mu\nu} \hat{R} - \frac{1}{\ell^2}
  g_{\mu\nu} = R_{\mu\nu} - \frac12 g_{\mu\nu} R - \frac{1}{\ell^2}
  g_{\mu\nu}  + [(n_\mu n_\nu) - g_{\mu\nu}] \Delta_\Sigma \delta_\Sigma,
\end{equation}
since the Ricci tensor \eqref{Ricci:con} and the curvature scalar
\eqref{R:con} for the conical space consist of the contributions from
the ordinary space and the conical defects.  So, any conical
singularity deforms the Einstein equation and it is responsible for
the off-shell solutions.  The off-shell free energy~\eqref{F:BH:off}
also follows a similar content in thermodynamics.  Thus, the off-shell
solutions are not the solutions of the Einstein equation, and they
will be used to interpolate the two on-shell states of the black hole
and the soliton.

In Fig.~\ref{fig:FM}, one can see that the end point ($M=0$) of the
free energy of the black hole is not connected with the free energy of
the soliton which is the thermal vacuum.  As is shown in section II,
the BTZ black hole is more probable below the critical temperature
while the thermal soliton is more probable above the critical
temperature.  To understand this Hawking-Page phase transition in
terms of the tunneling process, we have to obtain the continuous
off-shell free energy.  We expect there exist some missing off-shell
states in the region of the negative $M$.  So, we should consider
additional non-equilibrium states which can be characterized by the
conical space by allowing an arbitrary negative value of $M$.

For $M<0$, we now consider the space described by the
metric~\eqref{metric:euclidean} including a conical singularity at
$r=0$, which appears at the cone defined by the coordinate $r$ and
$\phi$.  The deficit angle and the area of $\Sigma$ are calculated as
\begin{align}
  \Delta_\Sigma &= 2\pi (1-\sqrt{-M}), \label{angle:deficit} \\
  A_\Sigma &= \int d\tau \sqrt{g_{\tau\tau}} = \beta
  \sqrt{-M}, \label{A:Sigma:sol}
\end{align}
when the period of the Euclidean time is $\beta$.  Then,
Eqs.~\eqref{I:G:E}, \eqref{I:GH:E}, and~\eqref{I:ct:E} are calculated
as
\begin{align}
  I_{\rm g} &= \frac{\beta}{4G} \left(\frac{r^2}{\ell^2} -M -
    \sqrt{-M} \right), \label{I:G:E:con} \\
  I_{\rm GH} &= \frac{\beta}{4G} \left(M - \frac{2r^2}{\ell^2}
  \right), \label{I:GH:E:con} \\
  I_{\rm ct} &= \frac{\beta}{4G} \frac{r}{\ell}
  \sqrt{f(r)}, \label{I:ct:E:con}
\end{align}
by taking into account  the conical space, which yields 
\begin{align}
  I_{\rm sol}^{\rm off} &= \frac{\beta}{4G} \left(\frac{r}{\ell}
    \sqrt{f(r)} - \frac{r^2}{\ell^2} - \sqrt{-M} \right)
  \notag \\
  &= \frac{\beta}{8G} \left[-M - 2\sqrt{-M} + O\left(\frac{1}{r^2}
    \right) \right]. \label{I:E:con}
\end{align}
Taking the infinite boundary, the off-shell free energy for the
soliton can be obtained as
\begin{align}
  F_{\rm sol}^{\rm off} = \frac{1}{8G} \left(-M - 2\sqrt{-M}
  \right). \label{F:con}
\end{align}
Note that it recovers the on-shell free energy~\eqref{F:AdS} for
$M=-1$. It is independent of the temperature in contrast to the case
of the black hole~\eqref{F:BH:off}.  In fact, the black hole can be
nucleated at the Hawking temperature; however, the soliton can be
formed at any temperature as long as $M=-1$.

\begin{figure}[htb]
  \centering
  \includegraphics[width=0.6\textwidth]{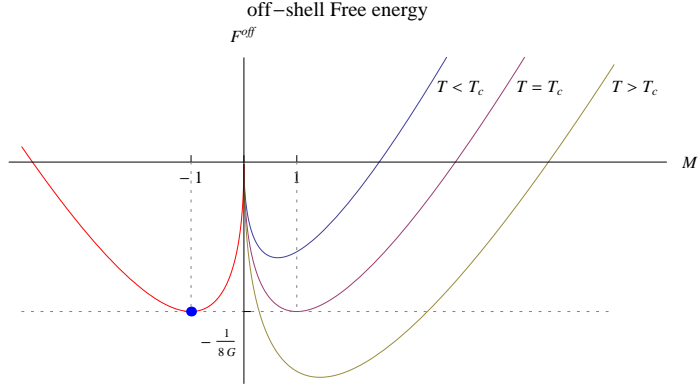}
  \caption{The three solid lines for $M>0$ show the off-shell free
    energies of the BTZ black hole depending on temperature of the
    heat reservoir, and the single solid line for $M<0$ shows the
    off-shell free energy of the soliton. The four minima consist of
    the three black hole states and one soliton state in each
    equilibrium. Note that $F^{\rm off}(0)=0$, which is finite.}
\label{fig:FvsMmod}
\end{figure}
The behavior of the off-shell free energy of the BTZ black hole and
the off-shell free energy of the soliton is shown in
Fig.~\ref{fig:FvsMmod}.  The off-shell free energy~\eqref{F:con} is
connected with the off-shell free energy of the BTZ black hole
continuously at $M=0$. Note that $F^{\rm off}$ is zero at $M=0$.  One
can easily see that the dot at $M=-1$ is just the thermal soliton of
the global AdS${}_3$ which is clearly stable.  Furthermore, from the
off-shell structure, it can be shown that the BTZ black hole decays
into the soliton below the critical temperature through the quantum
tunneling whereas the soliton can tunnel to the black hole above the
critical temperature.  The tunneling probability for this to occur
will be of the form~\cite{Hawking:1982dh}
\begin{align}
  \Gamma = A e^{-B}, \label{Gamma}
\end{align}
where $A$ is some determinant and $B$ is the absolute value of each
action of the BTZ black hole for $T<T_c$ or the soliton for $T>T_c$.

\section{Discussion}
\label{sec:discus}
We have studied the phase transition of the BTZ black hole in terms of
the off-shell free energy incorporated with the conical singularity
contributions.  In particular, thanks to the non-trivial off-shell
structure of the soliton, the isolated soliton state can be
continuously connected with the black hole state, so that the phase
transition which is associated with the tunneling effect can be easily
seen.

We have identified the off-shell free energy with a free energy in the
presence of the conical singularity whereas the on-shell free energy
is a free energy in the absence of the conical singularity.  In other
words, the criteria whether the free energy (or the action) is
on-shell or off-shell relies on the conical singularity
contributions. The reason for this is that if one defines an arbitrary
temperature, then it can cause conical singularities in the metric, so
that they affect the action through the modification of the curvature
scalar.  Actually, the nonrotating BTZ solution does not satisfy the
modified equation of motion given by Eq.~\eqref{eom:modified} when the
conical singularity in space is allowed.  Then, the system is in
off-shell which is corresponding to non-equilibrium state
thermodynamically.  Eventually, the off-shell free
energy~\eqref{F:BH:off} evaluated from the off-shell action by taking
into account the conical singularity turns out to be compatible with
the off-shell free energy~\eqref{Helmholtz} derived from the
conventional definition of the off-shell free energy.  Of course,
setting $\beta =\beta_{\rm H}$, the conical singularity can be removed
so that the system is in equilibrium. As for the soliton case of $M
<0$ which is a little bit different from the case of the black hole in
that there is no intrinsic temperature of the soliton, the temperature
of the soliton can be always the same with that of the heat
reservoir. In fact, there are two states at any temperatures, one is
the soliton in equilibrium satisfying the on-shell condition at
$M=-1$, and the others are in non-equilibrium.  In this respect, to
determine universally whether the system is in equilibrium or in
non-equilibrium, it is more plausible to justify the states in terms
of the conical singularity contributions to the free energy instead of
the temperature argument in our model.

Finally, we discuss two aspects on the cusp of the off-shell free
energy at $M=0$ which is not differentiable although it is
continuous. In particular, $F^{\rm off}(0)=0$ which is not singular
but finite as seen from Eqs.~\eqref{F:BH:off} and~\eqref{F:con} along
with Fig.~\ref{fig:FvsMmod}.  First, the cusp is a local maximum and
it seems to be a locally unstable state.  To clarify this point, let
us regard the off-shell free energy of $F^{\rm{off}}= E - TS$ as an
effective potential in finite temperature field theory as $Z=
e^{-\beta F^{\rm off}}= e^{{-\beta V_{\rm eff}}}$.  As is well-known,
the extrema of the effective potential are on-shell states which are
physical, while the other states are virtual.  Similarly, the extrema
from the off-shell free energy are regarded as equilibrium states
thermodynamically, whereas the others correspond to non-equilibrium
states because the only on-shell states satisfying $(dF^{\rm
  off}/dM)_T=0$ give the thermodynamic first law of $0=dF^{\rm{off}}=
dE - TdS $ where $T$ is the given temperature of the heat
reservoir. So, the on-shell states are physical from viewpoint of the
effective potential and they are also in equilibrium from
thermodynamical point of view. Now one can assume that some quantum
corrections may make the sharp free energy smooth, which yields a
smooth extremum. Eventually, we can regard this as a massless black
hole state mentioned in Ref.~\cite{Banados:1992wn}.  But it is
unstable as seen from the profile of the off-shell free energy in
Fig.~\ref{fig:FvsMmod}.  Note that we can not justify the stability of
the massless black hole in terms of the heat capacity~\eqref{C:BTZ}
since it vanishes at $M=0$.  So there are three on-shell states:
stable black hole, stable soliton, unstable massless black hole. From
the off-shell structure, we can see the massless black hole state
corresponding to the cusp is unstable.  Secondly, as for phase
transitions, the free energy is not singular but finite at $M =0$
since $F^{\rm off}(0)=0$, so that tunneling is possible obeying the
finite tunneling probability $\Gamma = A e^{-\beta |
  F|}$~\cite{Hawking:1982dh} where $F$ is the larger free energy among
stable states.  For instance, for the case of $T<T_c$, the BTZ black
hole undergoes a quantum tunneling and then decays thermodynamically
into the stable soliton whereas the soliton can tunnel to the black
hole for $T>T_c$.  This fact can be also understood from the analogous
role of the effective potential in finite temperature field theory. By
the way, one might ask whether any instanton solution connecting two
states is possible or not in connection with the tunneling effect.  We
can expect that it might be a BTZ type solution which asymptotically
goes to the BTZ solution at two end states.  Unfortunately, we do not
know what the explicit solution is, which is beyond the scope the
present work. We hope this interesting issue would be addressed
elsewhere.

\acknowledgments

W. Kim would like to thank E. J. Son for exciting discussions. M.~Eune
was supported by National Research Foundation of Korea Grant funded by
the Korean Government (Ministry of Education, Science and Technology)
(NRF-2010-359-C00007).  W.~Kim was supported by the Basic Science
Research Program through the National Research Foundation of
Korea(NRF) funded by the Ministry of Education, Science and
Technology(2012-0002880).  S.-H.Yi was supported by Basic Science
Research Program through the NRF of Korea funded by the
MEST(2012R1A1A2004410).  W.~Kim and S.-H.~Yi were in part supported by
the National Research Foundation of Korea(NRF) grant funded by the
Korea government(MEST) through the Center for Quantum
Spacetime(CQUeST) of Sogang University with grant number 2005-0049409.


\bibliographystyle{JHEP}
\bibliography{references}

\providecommand{\href}[2]{#2}\begingroup\raggedright\begin{thebibliography}{10}

\bibitem{Bekenstein:1973ur}
J.~D. Bekenstein, {\it {Black holes and entropy}},  {\em Phys.\ Rev.} {\bf D7}
  (1973) 2333--2346.

\bibitem{Gross:1982cv}
D.~Gross, M.~Perry, and L.~Yaffe, {\it {Instability of Flat Space at Finite
  Temperature}},  {\em Phys.\ Rev.} {\bf D25} (1982) 330--355.

\bibitem{Hawking:1982dh}
S.~Hawking and D.~N. Page, {\it {Thermodynamics of Black Holes in anti-De
  Sitter Space}},  {\em Commun.\ Math.\ Phys.} {\bf 87} (1983) 577.

\bibitem{York:1986it}
J.~York, James~W., {\it {Black hole thermodynamics and the Euclidean Einstein
  action}},  {\em Phys.\ Rev.} {\bf D33} (1986) 2092--2099.

\bibitem{Myers:1988ze}
R.~C. Myers and J.~Z. Simon, {\it {Black Hole Thermodynamics in Lovelock
  Gravity}},  {\em Phys.\ Rev.} {\bf D38} (1988) 2434--2444.

\bibitem{Banados:1992wn}
M.~Banados, C.~Teitelboim, and J.~Zanelli, {\it {The Black hole in
  three-dimensional space-time}},  {\em Phys.\ Rev.\ Lett.} {\bf 69} (1992)
  1849--1851, [\href{http://xxx.lanl.gov/abs/hep-th/9204099}{{\tt
  hep-th/9204099}}].

\bibitem{Mann:1990ci}
R.~B. Mann and T.~G. Steele, {\it {Thermodynamics and quantum aspects of black
  holes in (1+1)-dimensions}},  {\em Class.\ Quant.\ Grav.} {\bf 9} (1992)
  475--492.

\bibitem{Nappi:1992as}
C.~R. Nappi and A.~Pasquinucci, {\it {Thermodynamics of two-dimensional black
  holes}},  {\em Mod.\ Phys.\ Lett.} {\bf A7} (1992) 3337--3346,
  [\href{http://xxx.lanl.gov/abs/gr-qc/9208002}{{\tt gr-qc/9208002}}].

\bibitem{Zaslavsky:1994dx}
O.~Zaslavsky, {\it {Thermodynamics of (2+1) black holes}},  {\em Class.\
  Quant.\ Grav.} {\bf 11} (1994) L33--L38.

\bibitem{Reznik:1994py}
B.~Reznik, {\it {Thermodynamics and evaporation of the (2+1)-dimensions black
  hole}},  {\em Phys.\ Rev.} {\bf D51} (1995) 1728--1732,
  [\href{http://xxx.lanl.gov/abs/gr-qc/9403027}{{\tt gr-qc/9403027}}].

\bibitem{Peca:1998cs}
C.~S. Peca and P.~Lemos, Jose, {\it {Thermodynamics of Reissner-Nordstrom
  anti-de Sitter black holes in the grand canonical ensemble}},  {\em Phys.\
  Rev.} {\bf D59} (1999) 124007,
  [\href{http://xxx.lanl.gov/abs/gr-qc/9805004}{{\tt gr-qc/9805004}}].

\bibitem{Nojiri:1999pm}
S.~Nojiri and S.~D. Odintsov, {\it {Thermodynamics of Schwarzschild-(anti-)de
  Sitter black holes with account of quantum corrections}},  {\em Int.\ J.\
  Mod.\ Phys.} {\bf A15} (2000) 989--1010,
  [\href{http://xxx.lanl.gov/abs/hep-th/9905089}{{\tt hep-th/9905089}}].

\bibitem{Caldarelli:1999xj}
M.~M. Caldarelli, G.~Cognola, and D.~Klemm, {\it {Thermodynamics of
  Kerr-Newman-AdS black holes and conformal field theories}},  {\em Class.\
  Quant.\ Grav.} {\bf 17} (2000) 399--420,
  [\href{http://xxx.lanl.gov/abs/hep-th/9908022}{{\tt hep-th/9908022}}].

\bibitem{Medved:2000fk}
A.~Medved and G.~Kunstatter, {\it {One loop corrected thermodynamics of the
  extremal and nonextremal spinning BTZ black hole}},  {\em Phys.\ Rev.} {\bf
  D63} (2001) 104005, [\href{http://xxx.lanl.gov/abs/hep-th/0009050}{{\tt
  hep-th/0009050}}].

\bibitem{Nojiri:2001aj}
S.~Nojiri and S.~D. Odintsov, {\it {Anti-de Sitter black hole thermodynamics in
  higher derivative gravity and new confining deconfining phases in dual CFT}},
   {\em Phys.\ Lett.} {\bf B521} (2001) 87--95,
  [\href{http://xxx.lanl.gov/abs/hep-th/0109122}{{\tt hep-th/0109122}}].

\bibitem{Cvetic:2001bk}
M.~Cvetic, S.~Nojiri, and S.~D. Odintsov, {\it {Black hole thermodynamics and
  negative entropy in de Sitter and anti-de Sitter Einstein-Gauss-Bonnet
  gravity}},  {\em Nucl.\ Phys.} {\bf B628} (2002) 295--330,
  [\href{http://xxx.lanl.gov/abs/hep-th/0112045}{{\tt hep-th/0112045}}].

\bibitem{Cai:2006td}
R.-G. Cai, L.-M. Cao, and N.~Ohta, {\it {Mass and thermodynamics of
  Kaluza-Klein black holes with squashed horizons}},  {\em Phys.\ Lett.} {\bf
  B639} (2006) 354--361, [\href{http://xxx.lanl.gov/abs/hep-th/0603197}{{\tt
  hep-th/0603197}}].

\bibitem{Anninos:2008sj}
D.~Anninos and G.~Pastras, {\it {Thermodynamics of the Maxwell-Gauss-Bonnet
  anti-de Sitter Black Hole with Higher Derivative Gauge Corrections}},  {\em
  JHEP} {\bf 0907} (2009) 030, [\href{http://xxx.lanl.gov/abs/0807.3478}{{\tt
  arXiv:0807.3478}}].

\bibitem{Cadoni:2009bn}
M.~Cadoni and C.~Monni, {\it {BPS-like bound and thermodynamics of the charged
  BTZ black hole}},  {\em Phys.\ Rev.} {\bf D80} (2009) 024034,
  [\href{http://xxx.lanl.gov/abs/0905.3517}{{\tt arXiv:0905.3517}}].

\bibitem{Banerjee:2010sd}
R.~Banerjee and S.~Ghosh, {\it {Generalised Uncertainty Principle, Remnant Mass
  and Singularity Problem in Black Hole Thermodynamics}},  {\em Phys.\ Lett.}
  {\bf B688} (2010) 224--229, [\href{http://xxx.lanl.gov/abs/1002.2302}{{\tt
  arXiv:1002.2302}}].

\bibitem{Banerjee:2011au}
R.~Banerjee and D.~Roychowdhury, {\it {Thermodynamics of phase transition in
  higher dimensional AdS black holes}},  {\em JHEP} {\bf 1111} (2011) 004,
  [\href{http://xxx.lanl.gov/abs/1109.2433}{{\tt arXiv:1109.2433}}].

\bibitem{Horowitz:1998ha}
G.~T. Horowitz and R.~C. Myers, {\it {The AdS / CFT correspondence and a new
  positive energy conjecture for general relativity}},  {\em Phys.\ Rev.} {\bf
  D59} (1998) 026005, [\href{http://xxx.lanl.gov/abs/hep-th/9808079}{{\tt
  hep-th/9808079}}].

\bibitem{Surya:2001vj}
S.~Surya, K.~Schleich, and D.~M. Witt, {\it {Phase transitions for flat AdS
  black holes}},  {\em Phys.\ Rev.\ Lett.} {\bf 86} (2001) 5231--5234,
  [\href{http://xxx.lanl.gov/abs/hep-th/0101134}{{\tt hep-th/0101134}}].

\bibitem{Kleban:2004rx}
M.~Kleban, M.~Porrati, and R.~Rabadan, {\it {Poincare recurrences and
  topological diversity}},  {\em JHEP} {\bf 0410} (2004) 030,
  [\href{http://xxx.lanl.gov/abs/hep-th/0407192}{{\tt hep-th/0407192}}].

\bibitem{Carlip:1993sa}
S.~Carlip and C.~Teitelboim, {\it {The Off-shell black hole}},  {\em Class.\
  Quant.\ Grav.} {\bf 12} (1995) 1699--1704,
  [\href{http://xxx.lanl.gov/abs/gr-qc/9312002}{{\tt gr-qc/9312002}}].

\bibitem{Mann:1996ze}
R.~B. Mann and S.~N. Solodukhin, {\it {Quantum scalar field on
  three-dimensional (BTZ) black hole instanton: Heat kernel, effective action
  and thermodynamics}},  {\em Phys.\ Rev.} {\bf D55} (1997) 3622--3632,
  [\href{http://xxx.lanl.gov/abs/hep-th/9609085}{{\tt hep-th/9609085}}].

\bibitem{David:1999zb}
J.~R. David, G.~Mandal, S.~Vaidya, and S.~R. Wadia, {\it {Point mass
  geometries, spectral flow and AdS(3) - CFT(2) correspondence}},  {\em Nucl.\
  Phys.} {\bf B564} (2000) 128--141,
  [\href{http://xxx.lanl.gov/abs/hep-th/9906112}{{\tt hep-th/9906112}}].

\bibitem{Kurita:2004yn}
Y.~Kurita and M.-a. Sakagami, {\it {CFT description of three-dimensional
  Hawking Page transition}},  {\em Prog.\ Theor.\ Phys.} {\bf 113} (2005)
  1193--1213, [\href{http://xxx.lanl.gov/abs/hep-th/0403091}{{\tt
  hep-th/0403091}}].

\bibitem{Myung:2005ee}
Y.~S. Myung, {\it {No Hawking-Page phase transition in three dimensions}},
  {\em Phys.\ Lett.} {\bf B624} (2005) 297--303,
  [\href{http://xxx.lanl.gov/abs/hep-th/0506096}{{\tt hep-th/0506096}}].

\bibitem{Myung:2006sq}
Y.~S. Myung, {\it {Phase transition between the BTZ black hole and AdS space}},
   {\em Phys.\ Lett.} {\bf B638} (2006) 515--518,
  [\href{http://xxx.lanl.gov/abs/gr-qc/0603051}{{\tt gr-qc/0603051}}].

\bibitem{Maloney:2007ud}
A.~Maloney and E.~Witten, {\it {Quantum Gravity Partition Functions in Three
  Dimensions}},  {\em JHEP} {\bf 1002} (2010) 029,
  [\href{http://xxx.lanl.gov/abs/0712.0155}{{\tt arXiv:0712.0155}}].

\bibitem{Ogawa:2011fw}
N.~Ogawa and T.~Takayanagi, {\it {Higher Derivative Corrections to Holographic
  Entanglement Entropy for AdS Solitons}},  {\em JHEP} {\bf 1110} (2011) 147,
  [\href{http://xxx.lanl.gov/abs/1107.4363}{{\tt arXiv:1107.4363}}].

\bibitem{Grumiller:2012rt}
D.~Grumiller, R.~McNees, and S.~Zonetti, {\it {Black holes in the conical
  ensemble}},  \href{http://xxx.lanl.gov/abs/1210.6904}{{\tt arXiv:1210.6904}}.

\bibitem{Fursaev:1995ef}
D.~V. Fursaev and S.~N. Solodukhin, {\it {On the description of the Riemannian
  geometry in the presence of conical defects}},  {\em Phys.\ Rev.} {\bf D52}
  (1995) 2133--2143, [\href{http://xxx.lanl.gov/abs/hep-th/9501127}{{\tt
  hep-th/9501127}}].

\bibitem{Henningson:1998gx}
M.~Henningson and K.~Skenderis, {\it {The Holographic Weyl anomaly}},  {\em
  JHEP} {\bf 9807} (1998) 023,
  [\href{http://xxx.lanl.gov/abs/hep-th/9806087}{{\tt hep-th/9806087}}].

\bibitem{Balasubramanian:1999re}
V.~Balasubramanian and P.~Kraus, {\it {A Stress tensor for Anti-de Sitter
  gravity}},  {\em Commun.\ Math.\ Phys.} {\bf 208} (1999) 413--428,
  [\href{http://xxx.lanl.gov/abs/hep-th/9902121}{{\tt hep-th/9902121}}].

\bibitem{Hohm:2010jc}
O.~Hohm and E.~Tonni, {\it {A boundary stress tensor for higher-derivative
  gravity in AdS and Lifshitz backgrounds}},  {\em JHEP} {\bf 1004} (2010) 093,
  [\href{http://xxx.lanl.gov/abs/1001.3598}{{\tt arXiv:1001.3598}}].

\bibitem{Perez:2011qp}
A.~Perez, D.~Tempo, and R.~Troncoso, {\it {Gravitational solitons, hairy black
  holes and phase transitions in BHT massive gravity}},  {\em JHEP} {\bf 1107}
  (2011) 093, [\href{http://xxx.lanl.gov/abs/1106.4849}{{\tt
  arXiv:1106.4849}}].

\end{thebibliography}\endgroup


%
%

\end{document}